\documentclass{aa}
\usepackage{graphicx}
\usepackage{txfonts}
\newcommand{\mach}{${\mathcal M}$}

\begin{document}

   \title{Number ratios of young stellar objects in embedded clusters}

   \author{S. Schmeja
          \inst{1}
          \and
          R. S. Klessen\inst{1}
          \and
          D. Froebrich\inst{2}
          }

   \offprints{S. Schmeja, sschmeja@aip.de}

   \institute{Astrophysikalisches Institut Potsdam, An der Sternwarte 16,
              14482 Potsdam, Germany
%              \email{sschmeja;rklessen@aip.de}
         \and
             Dublin Institute for Advanced Studies, 5 Merrion Square, Dublin 2, Ireland
%             \email{df@cp.dias.ie}
             }

   \date{Received ; accepted }

   \abstract{
Embedded clusters usually contain young stellar objects in different
evolutionary stages. We investigate number ratios of objects in these classes
in the star-forming regions $\rho$~Ophiuchi, Serpens, Taurus, Chamaeleon~I,
NGC~7129, IC~1396A, and IC~348. They are compared to the temporal evolution of young
stars in numerical simulations of gravoturbulent fragmentation in order to
constrain the models and to possibly determine the evolutionary stage of the
clusters. It turns out that Serpens is the youngest, and IC~348 the
most evolved cluster, although the time when the observations are best
represented varies strongly depending on the model. Furthermore, we find an
inverse correlation of the star formation efficiency (SFE) of the models with
the Mach number. However, the observational SFE values cannot be reproduced by
the current isothermal models. This argues for models that take into account
protostellar feedback processes and/or the effects of magnetic fields.
   }

   \maketitle
%
%________________________________________________________________

\section{Introduction}

Almost all stars form in clusters. Embedded clusters contain various types of
young stars, making them ideal laboratories to study star formation as they
provide a large and genetically homogeneous sample (see Lada \& Lada
\cite{lada_lada} for a review).

Four classes of young stellar objects (YSOs) are distinguished
according to the properties of their spectral energy distributions
(SEDs). Originally Lada (\cite{1987IAUS..115....1L}) defined Class\,1,
2, and 3 objects according to the slope in the SED from 1 to
20\,$\mu$m. Note that Class\,2 sources correspond to classical
T\,Tauri stars and Class\,3 objects to weak line T\,Tauri stars. Later
the extremely embedded sources (Class\,0 objects) were added to this
classification, and their observational properties are defined e.g.\
in Andr\'e et al. (\cite{andre00}). These four classes are interpreted
as an evolutionary sequence from Class\,0 to 3. There are, however,
concerns that some Class\,0 sources might be mimicked by Class\,1
objects seen edge on (Men'shchikov \& Henning \cite{men_henn97}).
Furthermore there are works suggesting that Class\,2 and Class\,3
objects are actual of the same age (e.g. Walter
\cite{1986ApJ...306..573W}). Class\,0 sources are deeply embedded
protostars, possessing a large sub-mm ($\lambda >$\,350\,$\mu$m) to
bolometric luminosity ratio (L$_{\rm smm}$/L$_{\rm bol} >$\,0.005).
The Class\,0 stage is the main accretion phase and lasts only a few
10$^4$\,yr. Class\,1 objects are relatively evolved protostars. They
are surrounded by an accretion disc and a circumstellar envelope.
Pre-main-sequence stars in Class\,2 and 3 are characterised by a
circumstellar disc (optically thick in Class\,2, optically thin in
Class\,3) and the lack of a dense circumstellar envelope.  The
progenitors of these forming stars are prestellar cores (starless
cores, prestellar condensations). These are gravitationally bound,
dense molecular cloud cores with typical stellar masses that may
already be in a state of collapse, but have not formed a central
protostellar object yet.

Star formation in molecular clouds is controlled by the complex
interplay between interstellar turbulence and self-gravity
(V{\'a}zquez-Semadeni et al.\ \cite{VS_etal00}; Larson \cite{larson03};
Mac Low \& Klessen \cite{maclow_klessen04}, and references therein).
The supersonic turbulence ubiquitously observed in Galactic molecular
gas (Blitz \cite{blitz93}) generates strong density fluctuations with
gravity taking over in the densest and most massive regions (e.g.,
Sasao \cite{sasao73}; Hunter \& Fleck \cite{hunter_fleck82}; Elmegreen
\cite{elmegreen93}; Padoan \cite{padoan95}; Klessen \cite{klessen01};
Padoan \& Nordlund \cite{padoan_nordlund02}).  We call
this process gravoturbulent fragmentation.  In a cloud core where
gravitational attraction overwhelms all opposing forces from pressure
gradients or magnetic fields, localised collapse will set in. The
density increases until a protostellar objects forms in the centre and
grows in mass via accretion from the infalling envelope.  The
gravoturbulent models of molecular cloud evolution discussed here can
describe the entire collapse of a cloud core and the build-up of a
stellar cluster as a function of time (see, e.g., Klessen et al.\
\cite{KHM00}; Klessen \cite{klessen01}; Heitsch et al.\ \cite{HMK01};
Schmeja \& Klessen \cite{sk04}; Jappsen \& Klessen \cite{JK04}).
These models provide a ``snapshot'' of the
cluster at any time, allowing the comparison with observed clusters,
and possibly the determination of the cluster's evolutionary status
according to the models. This permits to constrain the models and
possibly to determine the evolutionary stage of the star-forming
region by comparison with the models.

Complete unbiased surveys of clusters for the content of all four
evolutionary classes and prestellar cores are difficult to undertake,
since they require different observational techniques. Hence,
different investigations have to be combined. The result therefore
might suffer from different detection limits or varying spatial
coverage. With this caveat in mind, we search the literature for
information on the different YSO classes in several star-forming
regions and tried to construct roughly homogeneous samples of Class\,0
to Class\,3 sources and prestellar cores to compare them with our
models.

The purpose of this paper is twofold: On the one hand, we compile
an observational sample of absolute numbers of YSOs
belonging to the different classes for several star-forming regions
from the literature (Sect.~\ref{sec:obsdata}), on the other hand we
analyse the evolution of the YSO classes in gravoturbulent models
(Sect.~\ref{sec:models}) and compare it
with the observational data in Sect.~\ref{sec:discussion}.
Finally, in Sect.~\ref{sec:conclude} we summarise and conclude our findings.

\section{Observational data}
\label{sec:obsdata}

The detection and classification of prestellar cores and
Class\,0/1/2/3 objects requires different observational techniques.
Thus, we have to construct our samples from various sources. We
consider corresponding areas on the sky, but the caveat remains that
the combined samples may be far from complete and not homogeneous.
Since it would hardly effect the relative numbers, the problem of
unresolved binaries can be neglected. Besides, we cannot resolve close
binaries in the SPH models either. The absolute and relative numbers
of YSOs adopted from the observations for the subsequent analysis are
listed in Table~\ref{tab:obs}. Prestellar cores are hard to determine
both from observations and in our models, and in particular they are
hard to compare, since the status of the cores (Jeans-critical or
subcritical) is often unknown.  Some cores considered as ``starless''
might even in fact turn out to harbour embedded sources (Young et al.\
\cite{young04}). Therefore we do not use them for the actual
comparison, but keep them as an additional test for consistency.
Due to constraints from the models (see Sect.~\ref{sec:models}) we
consider Class~2 and 3 combined.
So in the lower panel of Table~\ref{tab:obs} only Class~0, Class~1, and the
combined set of Class~2+3 objects are shown.

% \begin{table}[t]
%  \caption{Absolute {\em(upper panel)} and relative {\em(lower panel)}
%      numbers of YSOs of different classes from observations (references see text)}
%   \label{tab:obs}
%
% \begin{center}
% %\begin{minipage}{15cm}
% \begin{tabular}{l r r r r r l}
% \hline
% \hline
% Region & prestellar & 0 & 1 & 2 & 3 \\
% \hline
% $\rho$ Oph & 98 & 2 & 15 & 111 & 77 \\
% Serpens    & 26 & 5 & 19 & 22 & $\sim$20 \\
% Taurus     & 52 & 3 & 25 & 108 & 72 \\
% Cha I      & $<$71 & 2 & 5 & \multicolumn{2}{c}{175} \\
% NGC 7129   &    & 1 & 20 & 80  \\
% IC 1396A   &    & 2 & 6 & 47 & 1 \\
% \hline
% $\rho$ Oph &  & 0.01 & 0.07 & \multicolumn{2}{c}{0.92} \\
% Serpens    &  & 0.08 & 0.29 & \multicolumn{2}{c}{0.64} \\
% Taurus     &  & 0.01 & 0.12 & \multicolumn{2}{c}{0.87} \\
% Cha I      &  & 0.01 & 0.03 & \multicolumn{2}{c}{0.96} \\
% NGC 7129   &  & 0.01 & 0.20 & \multicolumn{2}{c}{0.79} \\
% IC 1396A   &  & 0.03 & 0.11 & \multicolumn{2}{c}{0.86} \\
% \hline
% \end{tabular}
% %\end{minipage}
% \end{center}
%  \end{table}

\subsection{$\rho$~Ophiuchi}

The \object{$\rho$~Ophiuchi molecular cloud} is the closest %($\sim$150~pc)
and probably best-studied star-forming region, offering the most
complete sample of YSOs.  Bontemps et al. (\cite{bontemps01}) find a
total number of 16 Class~1 sources, 123 Class~2 sources, 38 Class~3
sources, and 39 Class~3 candidates.  They did not detect the
previously known two Class~0 objects lying within their area.
Following the reasoning of Bontemps et al.\ (\cite{bontemps01}) and
the findings of Grosso et al.\ (\cite{grosso00}) that there might be
almost as many Class~3 as Class~2 objects, we add the candidates to
the Class~3 sample, yielding a total number of 77 Class~3 sources.
Stanke et al.\ (\cite{stanke04}) performed a 1.2\,mm dust continuum
survey of the $\rho$~Oph cloud and detected 118 starless clumps and a
couple of previously unknown protostars.
%About the same number of cores was detected by Johnstone et al.\ (\cite{johnstone04})
%by a 850~$\mu$m survey.
To obtain a reasonably homogeneous sample when combining the
Bontemps et al. (\cite{bontemps01}) and Stanke et al.\ (\cite{stanke04})
surveys
we restrict ourselves to the range covered by both investigations,
which is the $45' \times 45'$ region of the main $\rho$~Oph cloud
(\object{L1688}) investigated by Bontemps et al.\ (\cite{bontemps01}).
This area contains two Class~0 sources (Froebrich \cite{froebrich04};
Stanke et al.\ \cite{stanke04}).  The global star formation efficiency
(SFE) in L1688 is estimated at $6-14\%$, although the local SFE in the
subclusters where active star formation takes place is significantly
higher with $\sim 31\%$ (Bontemps et al.\ \cite{bontemps01}).  The
measured velocity dispersion is $2.6~\mathrm{km~s}^{-1}$ in
$\rho$~Oph~A and $2.7~\mathrm{km~s}^{-1}$ in $\rho$~Oph~B (Kamegai et
al.\ \cite{kamegai03}).  With the reported temperatures of 11 and
7.8~K this corresponds to Mach numbers of $\mathcal{M} \approx 13.5$
and $\mathcal{M} \approx 15.5$, respectively.

\begin{table}[t]
 \caption{Absolute {\em(upper panel)} and relative {\em(lower panel)}
     numbers of YSOs of different classes from observations (references see text)}
  \label{tab:obs}

\begin{center}
%\begin{minipage}{15cm}
\begin{tabular}{l r r r r r l}
\hline
\hline
Region & prestellar & 0 & 1 & 2 & 3 \\
\hline
$\rho$ Oph & 98 & 2 & 15 & 111 & 77 \\
Serpens    & 26 & 5 & 19 & 18 & $\sim$20 \\
Taurus     & 52 & 3 & 25 & 108 & 72 \\
Cha I      & $<$71 & 2 & 5 & \multicolumn{2}{c}{175} \\
IC 348     &    & 2 & 2  & \multicolumn{2}{c}{261} \\
NGC 7129   &    & 1 & 20 & 80  \\
IC 1396A   &    & 2 & 6 & 47 & 1 \\
\hline
$\rho$ Oph &  & 0.01 & 0.07 & \multicolumn{2}{c}{0.92} \\
Serpens    &  & 0.08 & 0.31 & \multicolumn{2}{c}{0.61} \\
Taurus     &  & 0.01 & 0.12 & \multicolumn{2}{c}{0.87} \\
Cha I      &  & 0.01 & 0.03 & \multicolumn{2}{c}{0.96} \\
IC 348     &  & 0.01 & 0.01 & \multicolumn{2}{c}{0.98} \\
NGC 7129   &  & 0.01 & 0.20 & \multicolumn{2}{c}{0.79} \\
IC 1396A   &  & 0.03 & 0.11 & \multicolumn{2}{c}{0.86} \\
\hline
\end{tabular}
%\end{minipage}
\end{center}
 \end{table}

\subsection{Serpens}

The \object{Serpens Cloud Core}, a very active, nearby star-forming region,
contains 26 probable protostellar condensations (Testi \& Sargent
\cite{testi_sargent}) and five Class~0 sources (Hurt \& Barsony
\cite{hurt_barsony}; Froebrich \cite{froebrich04}). Kaas et al.\
(\cite{kaas04}) detected 19 Class~1 and 18 Class~2 objects in the central
region (covering the field investigated by Testi \& Sargent
\cite{testi_sargent}). This is an unusually high Class~1/2 ratio compared to
other regions. Kaas et al.\ (\cite{kaas04}) cannot distinguish
between Class~3 sources and field stars and are therefore unable to give
numbers for Class~3. Preibisch (\cite{preibisch03}) performed an {\em XMM-Newton}
study of Serpens and detected 45 individual X-ray sources, most of them
Class~2 or Class~3 objects. Considering this and the argumentation of Kaas
et al.\ (\cite{kaas04}) we can assume that there are at least as many Class~3
as Class~2 sources in the relevant area. The local SFE in sub-clumps is around
9\%, the global SFE is estimated to be around $2-10\%$ (Kaas et al.\
\cite{kaas04}; Olmi \& Testi \cite{olmi_testi02}). The measured velocity
dispersion is $0.3 - 0.6~\mathrm{km~s}^{-1}$ (Olmi \& Testi
\cite{olmi_testi02}), corresponding to $1 \lesssim \mathcal{M} \lesssim 2.5$ at
$T = 20~\mathrm{K}$.

\subsection{Taurus}

The \object{Taurus molecular cloud} shows a low spatial density of
YSOs and represents a somewhat less clustered mode of low-mass star
formation. The numbers of Class~1, 2, and 3 sources in Taurus are
reported as 24, 108, and 72, respectively (Hartmann
\cite{hartmann02}). In the same area there are 52 prestellar cores
(Lee \& Myers \cite{lee_myers99}), and one Class~0 and three Class~0/1
sources (Froebrich \cite{froebrich04}).  For our analysis we divide
the Class~0/1 objects into two Class~0 and one Class~1 object.  Since
the prestellar cores in the sample of Lee \& Myers were selected by
optical extinction, their number may be underestimated compared to
other regions. Estimates for the star formation efficiency vary
between 2\% (Mizuno et al.\ \cite{mizuno95}) and 25\% in the dense
filaments (Hartmann \cite{hartmann02}). The velocity dispersion is
$0.49~\mathrm{km~s}^{-1}$ (Onishi et al. \cite{onishi96}),
corresponding to $\mathcal{M} \approx 2.5$, adopting a mean
temperature of $\sim 11$~K, i.e. a sound speed of
$0.2~\mathrm{km~s}^{-1}$.

%         Two column figure (place early!)
   \begin{figure*}
     \centering
     \includegraphics[width=17cm]{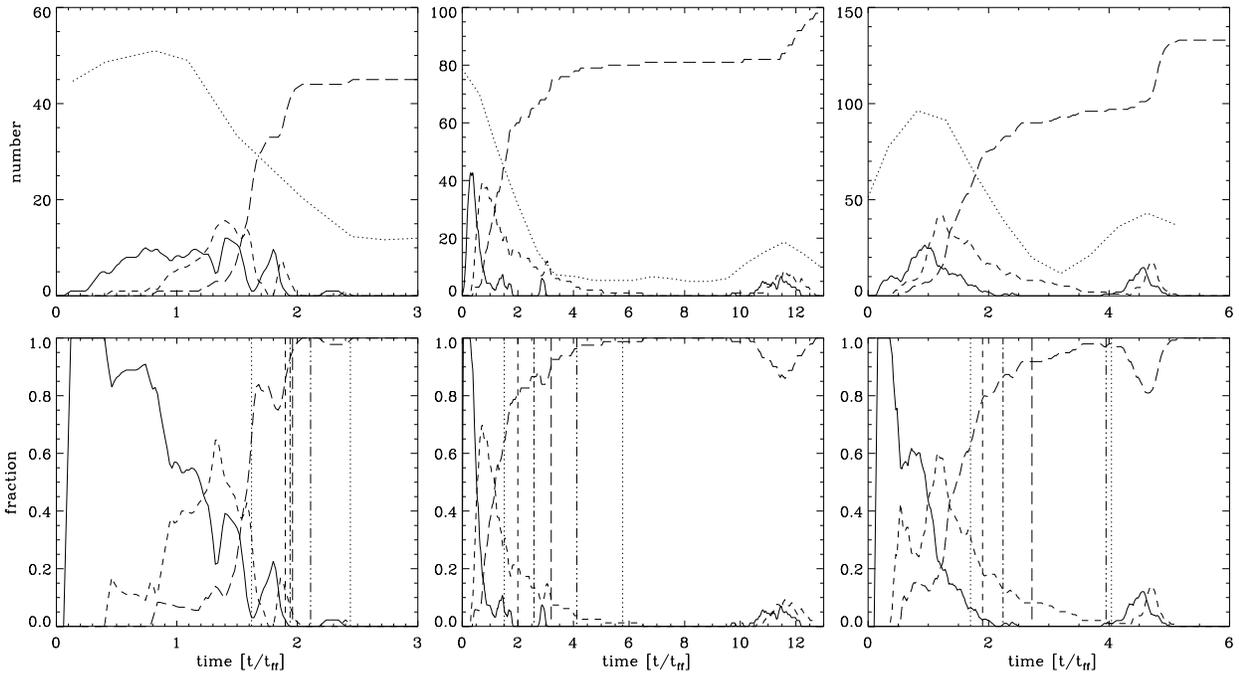}
     \caption{The temporal evolution of the fractions of YSO classes
       for three models: M2k4 (left), M6k2a (middle), and M10k2
       (right), {\em upper panel:} absolute numbers, {\em lower
         panel:} relative numbers.  Solid line: Class~0, dashed line:
       Class~1, long dashed line: Class~2+3, dotted line: prestellar
       cores (only shown in the upper panel).  The abscissa gives the
       time in units of the global free-fall time.  The zero point of
       the timescale corresponds to the time when gravity is
       ``switched on''.  Note the different intervals covered in the
       models.  The vertical lines in the lower panel indicate the
       time $\tau$, when the model shows the best agreement with the
       observations of Serpens (dotted), NGC~7129 (dashed),
       Taurus and IC~1396A (dash-dotted), $\rho$~Ophiuchi (long dashes),
       Cha~I (dash dot dot dot), and IC~348 (second dotted line).
       Note, because of similar relative
       YSO numbers the best-fit evolutionary times for Taurus and
       IC~1396A are essentially indistinguishable. }
               \label{fig:models}
    \end{figure*}

\subsection{Chamaeleon I}

The \object{Chamaeleon~I molecular cloud} harbours 126 confirmed and
54 new YSO candidates, most of them classical or weak-line T Tauri
stars (Class~2/3). Furthermore, four probable Class~1 protostars are
detected by the DENIS survey (Cambr\'esy et al.\ \cite{cambresy98}).
Persi et al.\ (\cite{persi01}) find two more Class~1 sources.  One
object is classified as Class~0, and one as Class~0/1 by Froebrich
(\cite{froebrich04}).  Counting the latter as Class~0 gives a total of
two Class~0 and five Class~1 sources.  Haikala et al.\
(\cite{haikala04}) detected 71 clumps (some of them associated with
embedded protostars) and a mean line width of
$0.62~\mathrm{km~s}^{-1}$ in the clumps, corresponding to $\mathcal{M}
\approx 3$ at $T \approx 11~\mathrm{K}$.  The SFE in Cha~I is about
13\% (Mizuno et al.\ \cite{mizuno99}).

\subsection{IC 348}

The young nearby cluster \object{IC 348} in the \object{Perseus molecular
cloud} complex contains 288 identified cluster
members (Luhman et al. \cite{luhman03}), including 23 brown dwarfs.
The majority of the objects are believed to be in the T~Tauri stage
(Class~2/3) of pre-main sequence evolution (Preibisch \& Zinnecker
\cite{pz04}). The active star formation phase seems to be finished
in the central parts of the cluster, but southwest of the cluster centre
a dense cloud core containing several embedded objects is found.
There are two Class~1 objects (Preibisch \& Zinnecker \cite{pz02}),
one confirmed Class~0 source and one Class 0 or 1 source (Froebrich
\cite{froebrich04}), which we count as Class~0 in our analysis.
Subtracting the brown dwarfs, we adopt a number of 261 Class~2/3,
two Class~1 and two Class~0 objects.
The average velocity dispersion is $1.04~\mathrm{km~s}^{-1}$
(Ridge et al.\ \cite{ridge03}), corresponding to $\mathcal{M}
\approx 5.2$ at $T \approx 11~\mathrm{K}$.

\subsection{NGC~7129 and IC~1396A}
\label{sec:ngc7129_ic1396}

With \object{NGC 7129} and \object{IC~1396A} we include two additional
star-forming regions in our analysis. Although, there is no
information on prestellar cores in the literature, sufficient data
about the population of young stellar objects can be found.
The numbers of YSOs in these two clusters are based on data from the
{\em Spitzer Space Telescope}, which are not available for the other regions.
Since {\em Spitzer} is very sensitive to the earliest YSO classes, the obtained
number ratios might be overestimated in favour of Class~0 and 1 compared
to the other regions.

In \object{NGC 7129} Muzerolle et al.\ (\cite{muzerolle04}) detected
one Class~0 object (classified as Class 0/1 by Froebrich
\cite{froebrich04}), 12 Class~1 objects and 18 Class~2 sources in
their {\em Spitzer} data.  They miss the core cluster members and estimate a
total of 20 Class~0/1 and 80 Class~2 objects, which is a similar ratio
as in Taurus or $\rho$~Oph. The average velocity dispersion is
$1.58~\mathrm{km~s}^{-1}$ (Ridge et al.\ \cite{ridge03}),
corresponding to $\mathcal{M} \approx 8$ at $T \approx 11~\mathrm{K}$.
This region has also been studied by Megeath et al.\
(\cite{megeath04}), see also below.

The Elephant Trunk Nebula, \object{IC~1396A}, was investigated by
Reach et al.\ (\cite{reach04}) using {\em Spitzer} data, revealing three
Class~0/1, five Class~1, 47 Class~2, and one Class~3 object.  These
numbers are similar to the numbers of YSOs found by Froebrich et al.\
(\cite{fsem04}), who however cannot distinguish between Class 1 and
Class 2/3 sources.  For our analysis we divide the Class~0/1 objects
into two Class~0 and one Class~1 object. Reach et al. (\cite{reach04})
estimate a SFE of $4-15\%$.

\subsection{Other star-forming regions}

Megeath et al.\ (\cite{megeath04}) report {\em Spitzer} results of the four
young stellar clusters \object{Cepheus~C}, \object{S171},
\object{S140}, and \object{NGC~7129}. They find ratios of Class~1 to
Class~2 objects between 0.37 and 0.57, which is significantly higher
than in the other regions listed in Table~\ref{tab:obs} except
Serpens. That indicates that these clusters are very young, although
the same caveat for {\em Spitzer} data as above applies.
Since no data of Class~3 objects are available we do not include these
clusters in our analysis.

%__________________________________________________________________

\section{The models}
\label{sec:models}

We perform numerical simulations of the fragmentation and collapse of
turbulent, self-gravitating gas clouds and the resulting formation and
evolution of protostars as described in Schmeja \& Klessen
(\cite{sk04}).  We use a code based on smoothed particle hydrodynamics
(SPH; Monaghan \cite{monaghan92}) in order to resolve large density
contrasts and to follow the evolution over a long timescale.  The code
includes periodic boundary conditions (Klessen \cite{klessen97}) and
sink particles (Bate et al.\ \cite{bbp95}) that replace high-density
cores while keeping track of mass and linear and angular momentum.
The periodic boundary conditions ensure that, independent of the box size,
all formed stars remain in the simulation, also later-type objects that
are known to be more widely distributed.
We determine the resolution limit of our SPH calculations using the Bate
\& Burkert (\cite{bate_burkert97}) criterion. This is sufficient for the
highly nonlinear density fluctuations created by supersonic turbulence as
confirmed by convergence studies with up to $10^7$ SPH particles. It
should be noted, however, that the Bate \& Burkert (\cite{bate_burkert97})
criterion may not be sufficient for describing the evolution of linear
perturbations close to equilibrium, as suggested by Klein et al.\
(\cite{KFM04}).

Our simulations consist of two globally unstable models that contract
from Gaussian initial conditions without turbulence and of 22 models
where turbulence is maintained with constant rms Mach numbers $\cal
M$, in the range $0.1 \le {\cal M} \le 10$.  We distinguish between
turbulence that carries its energy mostly on large scales, at
wavenumbers $1 \le k \le 2$, on intermediate scales, i.e.\ $3 \le k
\le 4$, and on small scales with $7 \le k \le 8$.  The naming of the
models, G1 and G2 for the Gaussian runs, and M\mach k$k$ (with rms
Mach number \mach\ and wavenumber $k$) for the turbulent models,
follows Schmeja \& Klessen (\cite{sk04}).  Details of the individual
models are given in their Table~1.

The dynamical behaviour of isothermal self-gravitating gas is scale free
and depends only on the ratio $\alpha$ between internal energy and potential energy:
$\alpha = E_{\rm int}/|E_{\rm pot}|$.
This scaling factor can be interpreted as a dimensionless temperature.
We convert to physical units by adopting a physical temperature of 11.3\,K
corresponding to an isothermal sound speed $c_\mathrm{s} = 0.2\,\mathrm{km\,s}^{-1}$,
and a mean molecular weight $\mu = 2.36$, corresponding to a typical value
in solar-metallicity Galactic molecular clouds.
With an average number density $n(\mathrm{H}_2) = 10^3\,$cm$^{-3}$, which
is consistent with the typical density in the considered star-forming regions,
the total mass in the two
Gaussian models is $2311\,\mathrm{M}_{\sun}$, and the size of the cube
is $3.4\,$pc.  The turbulent models have a mass of
$1275\,\mathrm{M}_{\sun}$ within a volume of ($2.8\,\mathrm{pc})^3$.
The global free-fall timescale is $\tau_{\rm ff} = 10^6\,$yr.
If we instead focus on individual dense cores like in
$\rho$~Oph with $n(\mathrm{H}_2) \approx 10^5\,$cm$^{-3}$,
the total masses in the Gaussian and in the turbulent models
are $231\,\mathrm{M}_{\sun}$ and $128\,\mathrm{M}_{\sun}$,
respectively, and the volumes are ($0.34\,\mathrm{pc})^3$ and
($0.28\,\mathrm{pc})^3$, respectively.
The global free-fall timescale is $\tau_{\rm ff} = 10^5\,$yr.
% and a mean density of  $n(\mathrm{H}_2) = 10^3\,$cm$^{-3}$, roughly
% corresponding to the mean density in the considered star-forming regions.
% Adopting a mean molecular weight $\mu = 2.36$, the total mass in the two
% Gaussian models is $xxx\,\mathrm{M}_{\sun}$, and the size of the cube
% is $x.x\,$pc.  The turbulent models have a mass of
% $xxx\,\mathrm{M}_{\sun}$ within a volume of ($x.x\,\mathrm{pc})^3$.
%The global free-fall timescale is $\tau_{\rm ff} = x^x\,$yr.
Note that the number of stars is not influenced by the adopted
physical scaling.
For further details on the scaling behaviour of the models see Klessen
\& Burkert (\cite{klessen_burkert00}) and Klessen et al.\ (\cite{KHM00}).

% The models are computed in normalised units.  Scaled to physical units
% we adopt a temperature of 11.3\,K corresponding to an isothermal sound
% speed $c_\mathrm{s} = 0.2\,\mathrm{km\,s}^{-1}$ and a mean density of
% $n(\mathrm{H}_2) = 10^5\,$cm$^{-3}$.  The total mass in the two
% Gaussian models is $220\,\mathrm{M}_{\sun}$, and the size of the cube
% is $0.34\,$pc.  The turbulent models have a mass of
% $120\,\mathrm{M}_{\sun}$ within a volume of ($0.28\,\mathrm{pc})^3$.
% The mean thermal Jeans mass in all models is $\langle M_{\rm J}
% \rangle = 1\,$M$_{\odot}$, and the global free-fall timescale is
% $\tau_{\rm ff} = 10^5\,$yr.
%%%%%%%%%

The YSO classes are determined as follows: The beginning of Class~0 is
considered the formation of the first hydrostatic core.  This happens
when the central object has a mass of about $0.01~\mathrm{M}_{\sun}$
(Larson \cite{larson03}).  The transition from Class~0 to Class~1 is
reached when the envelope mass is equal to the mass of the central
protostar (Andr\'e et al.\ \cite{andre00}).  The determination of the
end of the Class~1 stage is more difficult, since this is usually done
via spectral indices in the near-infrared part of the SED.  Generally,
after the Class~1 stage the objects are considered classical T~Tauri
stars that become visible in the optical.  Hence we determine the
transition from Class~1 to Class~2 when the optical depth of the
remaining envelope becomes unity at $2.2~\mu \mathrm{m}$ ($K$-band).
Using the evolutionary scheme of Smith (\cite{smith00}) and the
standard parameters as described in Froebrich et al.\ (\cite{fssk04})
the end of Class~0 corresponds to a mass of $M_* \approx
0.43~M_\mathrm{end}$, where $M_\mathrm{end}$ denotes the final mass of
the star. The end of Class~1 is reached when $M_* \approx
0.85~M_\mathrm{end}$. Note that the exact value of the mass at the
transition from one phase to the next does not influence our results
significantly.
Even a change of the opacity value by a factor of four results
in a deviation of the corresponding mass of a few per cent only.
Lacking a feasible criterion to distinguish Class~2 from Class~3 objects,
we consider both classes combined. The same is done for the observational
data.

Finding and defining prestellar cores is a more difficult task.
Usually one considers roughly spherical symmetrical density enhancements
containing no visible traces of protostars (e.g. Motte et al.\
\cite{motte98}; Johnstone et al.\ \cite{johnstone00}). We attempt to follow
this procedure and define prestellar cores from the Jeans-unstable
subset of all molecular cloud cores identified in our models. We use a
three-dimensional clump-finding algorithm to determine the cloud
structure, similar to the method of Williams et al.\
(\cite{williams94}). Further detail is given in Appendix A of Klessen
\& Burkert (\cite{klessen_burkert00}). This procedure matches many
of the observed structures and kinematic properties of nearby starless
molecular cloud cores (see the discussions in Ballesteros-Paredes et al.\
\cite{BKV03} and Klessen et al.\ \cite{KBVD04}).

%% RSK: modified...
%% The prestellar cores are
%% defined as Jeans-instable clumps, which probably match the observed
%% prestellar cores most closely.  They are detected using a
%% three-dimensional clump-finding algorithm similar to the method of
%% Williams et al.\ (\cite{williams94}), described in detail in Appendix
%% A of Klessen \& Burkert (\cite{klessen_burkert00}).

In order to avoid problems with the described mass criteria for the
determination of the classes for low-mass objects and to be consistent
with the observations, we only consider protostars with a final mass
$M_\mathrm{end} \ge 0.1~\mathrm{M}_{\sun}$, which roughly corresponds
to the detection limits of the observations reported in the
literature.  For the same reason we only consider models with a numerical
resolution of at least 200\,000 particles.  Furthermore, in order to
get reasonable numbers of protostars in the different classes we
select only those models where more than 37 protostars with
$M_\mathrm{end} \ge 0.1~\mathrm{M}_{\sun}$ are formed. This reduces
our set of models to 16. Again, see Table 1 of Schmeja \& Klessen
(\cite{sk04}) for further details.
%
% {\bf
% In some models, a small fraction of protostars gets highly
% accelerated (e.g. by ejection from a multiple system). These objects
% cross the box many times while they keep accreting. In reality,
% these protostars would have left the cluster and would not be able
% to gain any more mass. In these cases, we therefore consider accretion
% stopped after the object has crossed a volume corresponding to
% ten times the box size. The mass at this time is taken as the new
% final mass.
% }

%{\bf
In some models, a small fraction of protostars gets highly accelerated
(e.g.\ by ejection from a multiple system). Due to the adopted
periodic boundary conditions in our calculations, these objects cross
the computational domain many times while continuing to accrete. In
reality, however, these protostars would have quickly left the
high-density gas of the star-forming region and would not be able to
gain more mass. We therefore consider accretion to stop after the
object has crossed the computational box more than ten times.
Varying this distance does not significantly influence our
conclusions, e.g. increasing or decreasing it by a factor of two
changes the numbers derived in the next Section by less than 1\%.
%
% Making this distance larger or smaller by a factor of two does not
% influence our conclusion. The
%
% The mass at this time defines the final mass.
%}

%
%                                                One column figure
%----------------------------------------------------------- S_vib
   \begin{figure}
%   \centering
    \resizebox{\hsize}{!}{\includegraphics{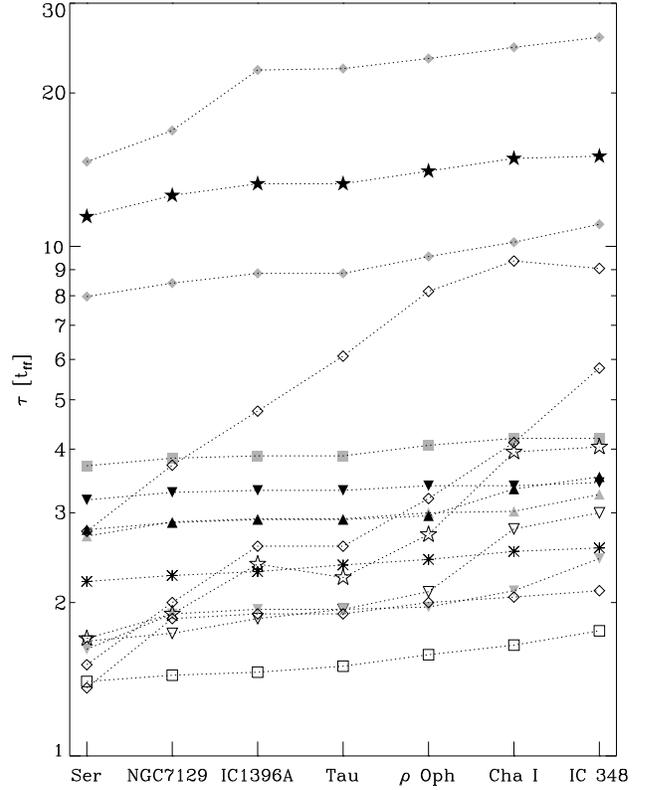}}
      \caption{The time $\tau$ of the best correspondence between models and
      observations (in units of the global free-fall time $\tau_\mathrm{ff}$)
         for the seven investigated clusters.
         The symbols denote the parameters of the models:
         %$\mathcal{M} = 0.1$ (circles; {\bf not shown}),
	 $\mathcal{M} = 0.5$ (triangles), $\mathcal{M} = 2$
         (triangles down), $\mathcal{M} = 3.2$ (squares), $\mathcal{M} = 6$ (diamonds),
         $\mathcal{M} = 10$ (stars); $1 \le k \le 2$ (open symbols),
         $3 \le k \le 4$ (grey symbols),  $7 \le k \le 8$ (filled symbols).
         The Gaussian collapse models are displayed as asterisks.
         %The dotted lines connect the symbols belonging together for the sake of clarity.
              }
         \label{fig:tau}
 \end{figure}

\section{Discussion}
\label{sec:discussion}

%\subsection{Temporal evolution of the models}
\subsection{The evolutionary sequence}

Figure~\ref{fig:models} shows the temporal evolution of the fractions
of the different YSO classes for three selected models. The formation
of the entire cluster takes place on varying timescales between about
two and 25 global free-fall times. In the Gaussian collapse models and
the turbulent models with small Mach numbers the formation tends to be
faster, because self-gravity dominates the large scales. The numbers
of Class~0 protostars show only a narrow peak, followed by a similar,
but shifted peak of Class~1 objects. In the models with higher
turbulence, the kinetic energy exceeds the gravitational one and the
system is formally supported on global scales. Collapse only occurs
locally in the shock compressed cloud clumps. The formation of Class~0
objects extends over a longer period (a few free-fall times) and in
some cases there is a second burst of star formation, following an
increase of the number of prestellar cores, at a later time as in
model M6k2a or M10k2. This second peak is not considered for the
comparison with the observations, though.

We count the numbers of objects in the particular classes and compare
the relative numbers to the observational values from
Table~\ref{tab:obs}.
We consider the entire observed population including more
dispersed objects. Due to the use of periodic boundary conditions
those are also included in the models.

The time $\tau$ of the best correspondence between
observations and models is determined when the weighted root mean
square of the differences
\begin{equation}
\sigma (t) = \sqrt{\frac{\frac{1}{2} \sum\limits_{i=0}^{2}{ \vert n_i^\mathrm{o} -
n_i^\mathrm{m}(t) \vert^2 \omega_i}} {\sum\limits_{i=0}^{2}
\omega_i}}
% \sigma (t) = \sqrt{\frac{1}{2} \sum\limits_{i=0}^{2}{ \vert n_i^\mathrm{o} -
% n_i^\mathrm{m}(t) \vert^2 \omega_i} / \sum\limits_{i=0}^{2}
% \omega_i}
\label{eq:sigma}
\end{equation}
becomes a minimum. The relative number of young stars in Class $i$ (0, 1, 2)
from observations is expressed as $n_i^{\rm o}$ and $n_i^{\rm m}$ denotes the
relative number of YSOs in Class $i$ from the models. The factor
$\omega_i$ is a weighting factor, introduced to account for the possible
scatter due to small number statistics in both, the observations and the
models. The weighting factor is set to $\omega_i = \sqrt{n_i^{\rm o} \cdot
n_i^{\rm m} }$.

   \begin{figure*}
   \centering
   \includegraphics[width=17cm]{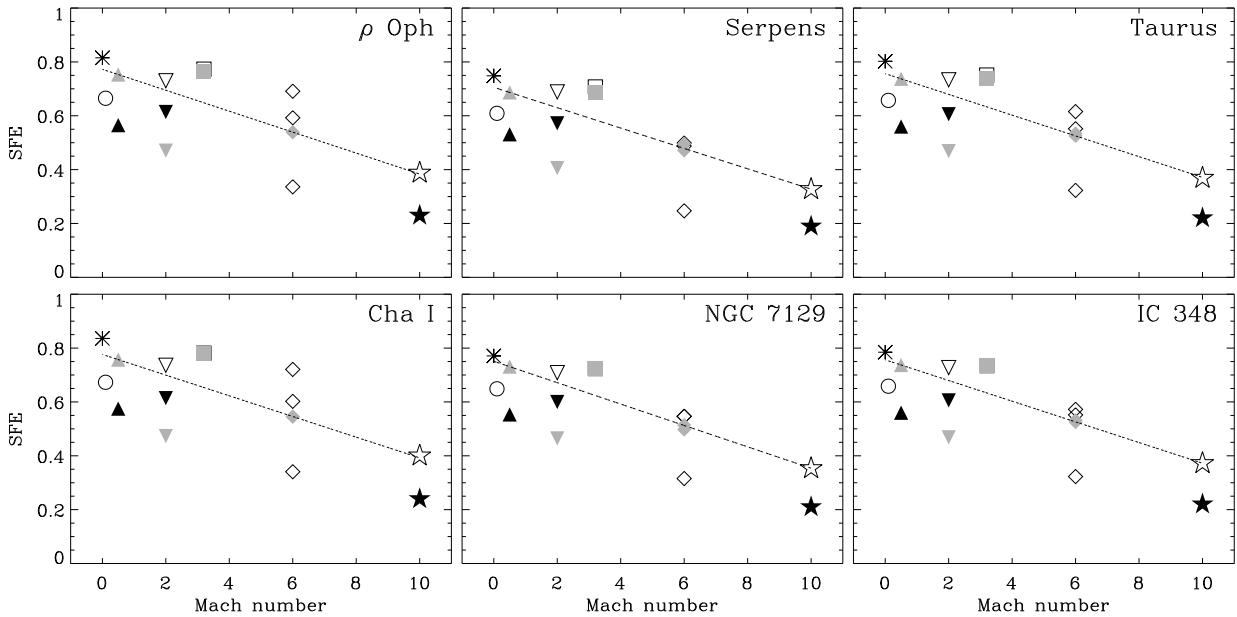}
     \caption{The star formation efficiency at time $\tau$ plotted versus the
       Mach number of the models.  The symbols are the same as in
       Fig.~\ref{fig:tau}, circles denote models with $\mathcal{M} = 0.1$.
       The dotted lines show a linear fit to the data.  }
               \label{fig:sfe}
    \end{figure*}

 The time $\tau$ is shown in the lower panel of Fig.~\ref{fig:models}
 and displayed for all models in Fig.~\ref{fig:tau}.  In general,
 Serpens is fitted worse than the other clusters. The mean minimal
 deviation of all models is
 $\langle \sigma_\mathrm{min} \rangle = 0.006$ for $\rho$~Oph, Cha~I, and IC~348,
 $\langle \sigma_\mathrm{min} \rangle = 0.008$ for Taurus,
 $\langle \sigma_\mathrm{min} \rangle = 0.009$ for IC~1396A,
 $\langle \sigma_\mathrm{min} \rangle = 0.012$  for NGC~7129, and
 $\langle \sigma_\mathrm{min} \rangle = 0.017$ for Serpens.
 In all cases $\sigma (t)$ has a clearly defined
 minimum.  The scatter of $\tau$ is high. Its value varies from about
 1.5 to 25 free-fall times for the different models, making it
 impossible to determine the actual ages of the clusters accurately.
 However, all models but one show the same sequence: $\tau$(Ser) $< \tau$(NGC
 7129) $< \tau$(IC 1396A) $\approx \tau$(Tau) $< \tau$($\rho$~Oph) $<
 \tau$(Cha~I) $< \tau$(IC~348).
The only exception is model M6k2c, where $\tau$(IC~348) is slightly
smaller than $\tau$(Cha~I).
 In five models the age of IC~1396A is smaller than that of Taurus,
in two models it is slightly larger, and in the other models the values are identical.
 Taking into account the second star-formation burst in the models
 M6k2a and M10k2 increases the ages of the older clusters, but it does
 not change the overall order.  Thus, independent of the applied
 model, Serpens is the youngest cluster, NGC~7129 the second youngest
 cluster, Cha~I and IC~348 are the most evolved
 clusters, while Taurus and IC~1396A are at roughly the same intermediate
 evolutionary stage.
%  The Taurus region, however, is generally not
%  described well by the gravoturbulent models (Froebrich et al.\
%  \cite{fssk04}), so the results on Taurus have to be taken with care.
However, the relative numbers of Class~0 and 1 objects in NGC~7129 and
IC~1396A may be overestimated due to possible observational biases
introduced by the {\em Spitzer Space Telescope}, as discussed in
Sect.~\ref{sec:ngc7129_ic1396}. These clusters may thus appear younger
relative to the other ones.
The fact that $\sigma (t)$ always has one well-defined minimum and that
all models produce the same evolutionary sequence independent of the
initial conditions, supports a scenario of a single ``burst''
of uninterrupted, rapid star formation. This is in agreement with other
theoretical and observational findings of star formation timescales and
molecular cloud lifetimes (see e.g. Hartmann et al.\ \cite{hbb01};
Ballesteros-Paredes \& Hartmann \cite {bh04}).
Several successive ``bursts'' of star formation within the cloud region
are likely to alter the picture and make a determination of the age from
the observed number ratios difficult.

\subsection{Star formation efficiency}
    
    Figure~\ref{fig:sfe} shows the star formation efficiency (the
    ratio of the mass accreted by the young stars to the total mass of
    the gas) at the time $\tau$ versus the Mach number of the models
    for six of the seven clusters.
    We find an inverse correlation of the SFE with the Mach number
    (false alarm probability $\le 0.5\%$). This is consistent with the
    theoretical findings by Klessen et al.\ (\cite{KHM00}) and Heitsch
    et al.\ (\cite{HMK01}), and probably due to the fact that in high-Mach
    number turbulence less mass is available for collapse at the sonic scale
    (V{\'a}zquez-Semadeni et al.\ \cite{VBK03}).  A linear fit is applied to
    the data and shown in Fig.~\ref{fig:sfe}.  There is no correlation
    with the driving wave number.  If we interpret the measured
    velocity dispersions in the clusters as the result of turbulence,
    we can estimate the SFEs at time $\tau$ for the particular Mach
    numbers from Fig.~\ref{fig:sfe}. In the case of $\rho$~Oph this
    requires to extrapolate the fit line beyond $\mathcal{M} = 10$.
    The SFE is $\sim 0.27$ in $\rho$~Oph,
    $\sim 0.45$ in NGC~7129, $\sim 0.56$ in IC~348,
    and between 0.60 and 0.65 in Serpens, Taurus, and Cha~I.
    (Lacking information on the velocity dispersion in IC~1396A, the
    corresponding SFE cannot be calculated.)
    These values
    are significantly higher than the measured SFEs, which are only around
    0.1.
    (No information is available on the SFEs of IC~348 and NGC~7129,
    but we expect them to be in the range of the other clusters.)
    Only in the case of $\rho$~Oph the SFE of the models is in
    the range of the SFE measured. The main reasons for this
    discrepancy are probably the limitation of the gas reservoir and
    the neglect of outflows and feedback mechanisms as well as of
    magnetic fields in the simulations.
    Bipolar outflows limit the local SFE, because the protostellar
    jet will carry a certain fraction of the infalling material away,
    furthermore, its energy and momentum input will affect the the protostellar
    envelope and may partially prevent it from accreting onto the protostar
    (e.g. Adams \& Fatuzzo \cite{af96}).
    The presence of magnetic fields would also retard the conversion of gas
    into stars (for current simulations see Heitsch et al.\ \cite{HMK01};
    V{\'a}zquez-Semadeni et al.\ \cite{vksb04}; Li \& Nakamura \cite{ln04}).

\subsection{Prestellar cores}

\begin{table}%[t]
 \caption{The ratio of prestellar cores to the total number of YSOs
     at time $\tau$ for three models and the observations}
  \label{tab:prestellar}
\begin{center}
%\begin{minipage}{15cm}
\begin{tabular}{l r r r r}
\hline
\hline
Region & M2k4 & M6k2a & M10k2 & Observations \\
\hline
$\rho$ Oph & 0.30 & 0.10 & 0.10 & 0.47 \\
Serpens & 0.79 & 0.06 & 0.74 & 0.39 \\
Taurus & 0.32 & 0.09 & 0.22 & 0.25 \\
Cha I & 0.18 & 0.11 & 0.82 & $<0.39$ \\
\hline
\end{tabular}
%\end{minipage}
\end{center}
\end{table}

As an additional test, the numbers of prestellar cores are analysed as
described in Sect.~\ref{sec:models} for the three models shown in
Fig.~\ref{fig:models} and compared to the four star-forming regions,
where information on the number of prestellar cores is available. The
ratios of the number of prestellar cores to the total number of YSOs
($n_\mathrm{psc}/n_*$) at time $\tau$ are listed in
Table~\ref{tab:prestellar} together with those values from the
observations (calculated from Table~\ref{tab:obs}). The observed ratio
in Taurus are roughly represented by models M2k4 and M10k2, but for
the other regions the models produce either a much higher or much
lower ratio. We intended to check if
$n_\mathrm{psc}^\mathrm{o}/n_\mathrm{psc}^\mathrm{m}$ allows us to
draw conclusions about the number of prestellar cores that actually
collapse and form stars. This seems not to be possible on the
basis of the current analysis.
In addition, defining a prestellar core is rather difficult,
both, from an observational and a theoretical point of view.
This adds another level of uncertainty to results based on
prestellar core statistics.

\section{Summary and conclusions}
\label{sec:conclude}

We analysed the temporal evolution of the fractions of YSO
classes in different gravoturbulent models of star formation and
compared it to observations of star-forming clusters. The observed
ratios of Class 0, 1, 2/3 objects in $\rho$~Ophiuchi, Serpens, Taurus,
Chamaeleon~I, NGC~7129, IC~1396A, and IC~348 can be reproduced by the
simulations, although the time when the observations are best
represented varies depending on the model. Nevertheless,
amongst the clusters with good observational sampling we always
find the following evolutionary sequence of increasing age:
Serpens, Taurus, $\rho$~Oph, Cha~I, and IC~348.

We find an inverse correlation of the star formation efficiency with
the Mach number. However, our models fail to reproduce the observed
SFEs for most of the clusters.  This is probably due to the lack of a
sufficiently large gas reservoir in the simulations and the neglect
of energy and momentum input from bipolar outflows and/or radiation
from young stars.  Only the SFE in $\rho$~Oph is reproduced. This
region is characterised by very high turbulent Mach numbers.  The fact
that our simple gravoturbulent models without feedback are able to
reproduce the right number ratios of YSOs for $\mathcal{M} \simeq 10$
suggests that protostellar feedback processes may not be important in
shaping the density and velocity structure in star-forming regions
with very strong turbulence and argues for driving mechanisms external
to the cloud itself (see also the discussion in Ossenkopf \& Mac~Low
\cite{Ossenkopf_MacLow02} and Mac~Low \& Klessen \cite{maclow_klessen04}).

The relative numbers of YSOs can reveal the evolutionary status of a
star-forming cluster only with respect to other clusters, the absolute
age is difficult to estimate. Better agreement between models and
observations requires a better consideration of environmental
conditions like protostellar outflows and magnetic fields in the
simulations. A larger observational sample, achieved by complete
censuses of more star-forming regions, is also desirable.

\begin{acknowledgements}
  We are very grateful to Roland Gredel, Thomas Stanke, Michael Smith
  and Tigran Khanzadyan for providing us with their $\rho$~Oph data
  prior to publication and to Lee Hartmann for sending us his data of
  the Taurus cloud.  We thank Michael Smith also for providing his
  evolutionary code.  The work of S.\,S. and R.\,S.\,K. is funded by
  the Emmy Noether Programme of the {\em Deutsche
    Forschungsgemeinschaft} (grant no.\ KL1358/1). D.\,F. received
  financial support by the {\em Cosmo-Grid} project, funded by the
  Program for Research in Third Level Institutions under the National
  Development Plan and with assistance from the European Regional
  Development Fund.  This publication makes use of the Protostars
  Webpage ({\tt www.dias.ie/protostars/}) hosted by the Dublin
    Institute for Advanced Studies.

\end{acknowledgements}


\begin{thebibliography}{}


%%%%%%%%%%%%%%%%%%%%%%%%%%%%%%%%%%%%%

\bibitem[1996]{af96} Adams, F.~C., \& Fatuzzo, M. 1996, \apj, 464, 256

\bibitem[2000]{andre00} Andr\'{e}, P., Ward-Thompson, D., \& Barsony, M. 2000, in
       Proto\-stars and Planets IV, ed.\ V. Mannings, A.~P. Boss, \& S.~S. Russell
       (Tucson: University of Arizona Press), 59

\bibitem[2005]{bh04} Ballesteros-Paredes, J., \& Hartmann, L. 2005, ApJ, submitted

\bibitem[2003]{BKV03} Ballesteros-Paredes, J., Klessen, R.\ S., \&
  V{\'a}zquez-Semadeni, E. 2003, \apj, 592, 188

\bibitem[1997]{bate_burkert97} Bate, M.\ R., \& Burkert, A. 1997, \mnras, 288, 1060

\bibitem[1995]{bbp95} Bate, M.~R., Bonnell, I.~A., \& Price, N.~M. 1995, \mnras, 277, 362

\bibitem[1993]{blitz93} Blitz, L. 1993, in Protostars and Planets III,
eds. E.~H. Levy \& J.~I. Lunine (Tucson: Univ. Arizona Press), 125

\bibitem[2001]{bontemps01} Bontemps, S., Andr\'e, P., Kaas, A.~A., et al.\ 2001,
        \aap, 372, 173

\bibitem[1998]{cambresy98} Cambr\'esy, L., Copet, E., Epchtein, N., et al.\ 1998, \aap, 338, 977

\bibitem[1993]{elmegreen93} Elmegreen, B.~G.\ 1993, \apjl, 419, L29

% \bibitem[2004]{FKM04} Fisher, R.\ T., Klein, R.\ I., \& McKee, C.\
%   F. 2004, in preparation

\bibitem[2005]{froebrich04} Froebrich, D. 2005, \apjs, 156, 169

\bibitem[2005a]{fssk04} Froebrich, D., Schmeja, S., Smith, M.~D., \& Klessen, R.~S.
        2005a, \aap, submitted

\bibitem[2005b]{fsem04} Froebrich, D., Scholz, A., Eisl\"offel, J., \& Murphy, G.~C. 2005b,
        \aap, 432, 575

\bibitem[2000]{grosso00} Grosso, N., Montmerle, T., Bontemps, S., Andr\'e, P.,
        \& Feigelson, E.~D. 2000, \aap, 359, 113

\bibitem[2005]{haikala04} Haikala, L.~K., Harju, J., Mattila, K., \& Toriseva, M.
2005, \aap, 431, 149

\bibitem[2002]{hartmann02} Hartmann, L. 2002, \apj, 578, 914

\bibitem[2001]{hbb01} Hartmann, L., Ballesteros-Paredes, J., \& Bergin, E.~A.
     2001, \apj, 562, 852

\bibitem[2001]{HMK01} Heitsch, F., Mac Low, M.-M., \& Klessen, R.\ S.\ 2001, \apj, 547,  280

\bibitem[1982]{hunter_fleck82} Hunter, J.~H., \& Fleck, R.~C.\ 1982, \apj, 256, 505

\bibitem[1996]{hurt_barsony} Hurt, R.~L., \& Barsony, M. 1996, \apj, 460, L45

\bibitem[2004]{JK04} Jappsen, A.-K., \& Klessen, R.\ S. 2004, \aap,
  423, 1

\bibitem[2000]{johnstone00} Johnstone, D., Wilson, C.~D., Moriarty-Schieven, G. et al. 2000, \apj, 545, 327

% \bibitem[2004]{johnstone04} Johnstone, D., Di Francesco, J., \& Kirk, H. 2004, ApJ, 611, L45

\bibitem[2004]{kaas04} Kaas, A.~A., Olofsson, G., Bontemps, S., et al.\ 2004, \aap, 421, 623

\bibitem[2003]{kamegai03} Kamegai, K., Ikeda, M., Maezawa, H., et al. 2003, \apj, 589, 378

\bibitem[2004]{KFM04} Klein, R.~I., Fisher, R., \& McKee, C.~F. 2004,
%in Gravitational Collapse: From Massive Stars to
 %Planets, eds.\ G.~Garc{\'{\i}}a-Segura, G.~Tenorio-Tagle, J. Franco, \& H.\,W. Yorke,
 Rev.\ Mex.\ Astron.\ Astrof{\'{\i}}s.\ (Ser.\ de Conf.), 22, 3

\bibitem[1997]{klessen97} Klessen, R.~S. 1997, \mnras, 292, 11

\bibitem[2001]{klessen01} Klessen, R.~S. 2001, \apj, 556, 837

\bibitem[2000]{klessen_burkert00} Klessen, R.~S., \& Burkert, A. 2000, \apjs, 128, 287

\bibitem[2000]{KHM00} Klessen, R.~S., Heitsch, F., \& Mac Low, M.-M.,\ 2000, \apj, 535, 887

\bibitem[2005]{KBVD04} Klessen, R.~S., Ballesteros-Paredes, J., V{\'a}zquez-Semadeni, E.,
\& Dur{\'a}n-Rojas, C. 2005, \apj, 620,786

\bibitem[1987]{1987IAUS..115....1L}
Lada, C.~J. 1987, in Star Forming Regions, IAU Symp. 115, 1
%Star formation - From OB associations to protostars

\bibitem[2003]{lada_lada} Lada, C.~J., \& Lada, E.~A. 2003, \araa, 41, 57

\bibitem[2003]{larson03} Larson, R.~B. 2003, Rep.\ Prog.\ Phys., 66, 1651

\bibitem[1999]{lee_myers99} Lee, C.~W., \& Myers, P.~C. 1999, \apjs, 123, 233

\bibitem[2004]{ln04} Li, Z.-Y., \& Nakamura, F. 2004, \apj, 609, L83

\bibitem[2003]{luhman03} Luhman, K.~L., Stauffer, J.~R., Muench, A.~A., et al.
2003, \apj, 593, 1093

\bibitem[2004]{maclow_klessen04} Mac Low, M.-M., \& Klessen, R.~S. 2004, Rev.\ Mod.\ Phys., 76, 125

\bibitem[2004]{megeath04} Megeath, S.~T., Allen, L.~E., Gutermuth,
  R.~A., et al. 2004, \apjs, 154, 367

\bibitem[1997]{men_henn97} Men'shchikov, A. B., \& Henning, T. 1997, \aap, 318, 879

\bibitem[1995]{mizuno95} Mizuno, A., Onishi, T., Yonekura, Y., et al.\ 1995, \apj, 445, L161

\bibitem[1999]{mizuno99} Mizuno, A., Hayakawa, T., Tachihara, K., et al.\ 1999, \pasj, 51, 859

\bibitem[1992]{monaghan92} Monaghan, J.~J. 1992, \araa, 30, 543

\bibitem[1998]{motte98} Motte, F., \& Andr\'e, P., \& Neri, R. 1998, \aap, 336, 150

\bibitem[2004]{muzerolle04} Muzerolle, J., Megeath, S.~T., Gutermuth, R.~A., et al.\ 2004,
\apjs, 154, 379

\bibitem[2002]{olmi_testi02} Olmi, L., \& Testi, L. 2002, \aap, 392, 1053

\bibitem[1996]{onishi96} Onishi, T., Mizuno, A., Kawamura, A., Ogawa, H., \& Fukui, Y.
      1996, \apj, 465, 815

\bibitem[2002]{Ossenkopf_MacLow02} Ossenkopf, V., \& Mac Low, M.-M. 2002, \aap, 390, 307

\bibitem[1995]{padoan95} Padoan, P.\ 1995, \mnras, 277, 377

\bibitem[2002]{padoan_nordlund02} Padoan, P., \& Nordlund, {\AA}. 2002, \apj, 576, 870

\bibitem[2001]{persi01} Persi, P., Marenzi, A.~R., G\'omez, M., \& Olofsson, G. 2001,
\aap, 376, 907

\bibitem[2003]{preibisch03} Preibisch, T. 2003, \aap, 410, 951

\bibitem[2002]{pz02} Preibisch, T., \& Zinnecker, H. 2002, \aj, 123, 1613

\bibitem[2004]{pz04} Preibisch, T., \& Zinnecker, H. 2004, \aap, 422, 1001

\bibitem[2004]{reach04} Reach, W.~T., Rho, J., Young, E., et al.\
  2004, \apjs, 154, 385

\bibitem[2003]{ridge03} Ridge, N.~A., Wilson, T.~L., Megeath, S.~T., Allen, L.~E., \& Myers, P.~C.
   2003, \aj, 126, 286

\bibitem[1973]{sasao73} Sasao, T. 1973, \pasj, 25, 1

\bibitem[2004]{sk04} Schmeja, S., \& Klessen, R.~S. 2004, \aap, 419, 405

\bibitem[2000]{smith00} Smith, M.~D. 2000, Ir.\ Astron. J., 27, 25

\bibitem[2004]{stanke04} Stanke, T., Smith, M.~D., Gredel, R., \& Khanzadyan, T. 2004, \aap,
        submitted

\bibitem[1998]{testi_sargent} Testi, L., \& Sargent, A.~I. 1998, \apj, 508, L91

\bibitem[1986]{1986ApJ...306..573W}
Walter, F.~M. 1986, \apj, 306, 573
%X-ray sources in regions of star formation. I - The naked T Tauri stars

\bibitem[2000]{VS_etal00} V\'azquez-Semadeni, E., Ostriker, E. C., Passot, T., Gammie, C.
\& Stone, J. 2000, in Protostars and Planets IV, ed.\ V. Mannings, A.~P. Boss,
\& S.~S. Russell (Tucson: University of Arizona Press), 3

\bibitem[2003] {VBK03} V{\' a}zquez-Semadeni, E.,
  Ballesteros-Paredes, J., \& Klessen, R.~S.\ 2003, \apjl, 585, L131

\bibitem[2005] {vksb04} V{\' a}zquez-Semadeni, E., Kim, J., Shadmehri, M., \& Ballesteros-Paredes,
J. 2005, \apj, 618, 344

\bibitem[1994]{williams94} Williams, J.~P., De Geus, E.~J., \& Blitz, L. 1994, \apj, 428, 693

\bibitem[2004]{young04} Young, C.~H., J{\o}rgensen, J.~K., Shirley,
    Y.~L., et al. 2004, \apjs, 154, 396

\end{thebibliography}
\end{document}